\documentclass{mn2e}
\usepackage{epsf}
\usepackage{times}

\begin{document}

\title{Improved approximations of Poissonian errors for high confidence levels}
\author[Harald Ebeling]{Harald Ebeling\\
Institute for Astronomy, University of Hawaii, 2680 Woodlawn Drive,
Honolulu, HI 96822, USA}

\maketitle

\begin{abstract}
We present improved numerical approximations to the exact Poissonian
confidence limits for small numbers $n$ of observed events following
the approach of Gehrels (1986). Analytic descriptions of all
parameters used in the approximations are provided to allow their
straightforward inclusion in computer algorithms for processing of
large data sets. Our estimates of the upper (lower) Poisson confidence
limits are accurate to better than 1\% for $n\le 100$ and values of
$S$, the derived significance in units of Gaussian standard
deviations, of up to 7 (5). In view of the slow convergence of the
commonly used Gaussian approximations toward the correct Poissonian
values, in particular for higher values of $S$, we argue that, for
$n\le 40$, Poissonian statistics should be used in most applications,
unless errors of the order of, or exceeding, 10\% are acceptable.
\end{abstract}


\section{Introduction}

The need to assess the statistical significance of an observed small
number of events is common in astrophysics as well as in virtually all
other natural sciences. The number of neutrinos detected in an
underground detector, the number of supernovae observed at $z>1$, the
number of photons in a faint X-ray point source --- all of these are
numbers likely to be in the Poisson regime, and knowing the correct
errors of these measurements is obviously crucial to any scientific
conclusions drawn from these numbers.

In a much noted paper Gehrels (1986, from here on G86) provided
analytic approximations to the correct Poisson confidence limits for
small event numbers. A particulary useful new and improved
approximation in G86 is for the Poisson lower confidence limit,
$\lambda_l$ (see Section~\ref{nom} below for an overview of the
nomenclature used). However, two issues limit the applicability of
these approximations to large data sets and/or applications that
require high confidence levels. First, Gehrels' analytic approximation
to the correct value of $\lambda_l$ uses empirically determined
parameters $\beta$ and $\gamma$, both of which are non-trivial
functions of $S$, the desired significance in units of Gaussian
standard deviations.  G86 tabulates $\beta(S)$ and $\gamma(S)$ for ten
values of $S$ ranging from one to 3.3$\sigma$ but does not provide an
analytic formula that would allow the reader to compute $\lambda_l$
for arbitrary $S$ values\footnote{Especially relevant in view of the
fact that $\gamma(S)$ actually becomes singular close to $S=1$, as we
show in Section~\ref{lowlim}.}. Second, the validity of all
approximations presented and discussed in G86 has only been verified
in the same, relatively narrow range of confidence levels from 1 to
$3.3\sigma$. To our knowledge, no reasonably accurate approximations
have been published for $S>3.3\sigma$.

While these issues may be of limited importance for many, if not most,
applications, they can become serious in cases where confidence limits
need to be computed for large sets of event numbers, particularly if
high confidence levels are required. Consider, for instance, X-ray
astronomy, a traditionally photon starved line of observational
research. With large-area, high-resolution X-ray CCD detectors now in
use on board the Chandra and XMM-Newton X-ray Observatories, X-ray
images of dimensions $1024\times 1024$ or even $4096\times 4096$
pixels have become common. The numbers of photons registered in the
vast majority of these pixels will be in the Poisson regime, and
assessing accurately the significance of any features embedded in the
very low and spatially non-uniform background measured with these
large arrays requires the accurate computation of a considerable
number of Poisson confidence limits. A real-life example of a
scientific project relevant in this context is the compilation of a
statistical sample of unresolved (single-pixel) point sources detected
at the greater than $5\sigma$ confidence level (i.e., $S=5$) in a set
of Chandra ACIS-I images.

It is for applications like the ones outlined above that we here
present modified and improved versions of G86's approximations that
are more accurate over an extended range of confidence levels ($S\la
7\sigma$). To permit these approximations to be incorporated
straightforwardly into computer algorithms for the processing of large
data sets, we also provide analytic descriptions of all parameters
used, thus allowing the computation of accurate Poisson confidence
limits for a wide range of event numbers and confidence levels.

\section{Nomenclature}
\label{nom}

In the following we adopt Gehrels' nomenclature and definitions.
Specifically, upper limits $\lambda_u$ and lower limits $\lambda_l$
are, for Poisson statistics, defined by
\begin{equation}
\sum_{x=0}^{n}\; \frac{\lambda_u^n\; {\rm e}^{-\lambda_u}}{x!} = 1 - {\rm CL}
\hspace*{1cm} n\ge0 
\end{equation}
and
\begin{equation}
\sum_{x=0}^{n-1}\; \frac{\lambda_l^n\; {\rm e}^{-\lambda_l}}{x!} = {\rm CL}
\hspace*{1cm} n\ge1 
\end{equation}
where $n$ is the number of events observed, and CL is the desired
confidence level (Eq.~1 and 2 of G86). (For $n=$ 0, $\lambda_l=$0 for
all values of CL). For Gaussian statistics (i.e.\ a probability
distribution which is normal) CL is related to $S$, the equivalent
Gaussian number of $\sigma$, by
\begin{equation}
{\rm CL}(S) = \frac{1}{\sqrt{2\pi}}\; \int\limits_{-\infty}^{S} {\rm
e}^{-t^2/2}\; dt\,.
\end{equation}
For ease of presentation we shall use $S$ to parametrize a large range
of CL values.

For $n=0$ to 100 and selected values of $S$ ranging from 1 to 3.3 G86
tabulates $\lambda_u$ and $\lambda_l$ as obtained from Eqs.~1 and 2;
Gehrels also presents analytic and numerical approximations accurate
to better then 2\% for $S<3.3$. In this paper, we test the Gaussian
approximation as well as the ones discussed in G86 over a larger range
of confidence levels ($S\le 7$). We then modify Gehrels approximations to
improve their performance specifically for large values of $S$ and,
finally, present numerical (polynomial) descriptions of all parameters
used, to allow the computation of approximate values of
$\lambda_u\,(n,S)$ and $\lambda_l\,(n,S)$ for all values of $n$ up to
at least 100 and $S\le 7$.

\section{The Gaussian approximation}

For a normal probability distribution (i.e., in the limit
$n\rightarrow \infty$) $\lambda_u$ and $\lambda_l$ are given by the
well known expressions $n \pm S\sqrt{n}$ which are also commonly used
to approximate the not straightforwardly computable Poisson limits for
`reasonably' large values of $n$. What `reasonable' means in this
context is subject to debate; in practice, values of $n$ in excess of
20 (and sometimes even $n>10$) are often deemed sufficiently large to
justify the use of the Gaussian approximation.

In Figure~1, we show the percentage error of the Gaussian
approximation $n \pm S\sqrt{n}$ when applied to event numbers in the
Poisson regime (loosely defined as $n<100$). For low to moderate
confidence levels ($S\le 2$) the Gaussian approximation is accurate to
better than 10\% for $n=20$ (but not for $n=10$!), which may or may
not be sufficient for a given application. However, the error of the
Gaussian approximation, in particular for the lower limit, increases
rapidly for higher confidence levels. Already at $S=3$, $n>45$ is
required to limit the error in $\lambda_l$ to 10\%; for $S=5$ even
$n=100$ is insufficient if 10\% accuracy are sought. 10\% accuracy may
not nearly be good enough though in applications where errors of many
independent measurements are propagated. In the case of X-ray spectral
fitting, for instance, theoretical models are fitted simultaneously to
events registered in hundreds of independent energy channels. A {\em
systematic}\/ error of the order of 10\% in the errors on the counts
in {\em all}\/ spectral bins introduced by the use of Gaussian
approximations to the true Poisson errors can lead to best-fit
parameter values that may be erroneous by significanly more than the
formal, statistical errors.

\section{Approximation of the Poissonian upper limit}

G86 derives two analytic approximations to the true Poissonian upper
limit, namely
\[
 \lambda_u \approx (n+1)\; \left[ 1 - \frac{1}{9\,(n+1)} +
                   \frac{S}{3\,\sqrt{n+1}} \right]^3 
  \hspace*{1cm} \mbox{(G86 Eq. 9)} 
\]
and
\[
 \lambda_u \approx n + S\,\sqrt{n+1} + \frac{S^2+2}{3}
  \hspace*{1cm} \mbox{(G86 Eq. 10)}
\]
where the latter expression is simply the previous one expanded and
limited to the dominant terms in $(n+1)$.

In Figure~2 we show the percentage error of these two approximations
for a range of $n$ and $S$ values. Both expressions represent a
considerable improvement over the Gaussian approximations (cf.\
Fig.~1). At low values of $n$ the remaining error of several per cent
may, however, still be too high for certain applications.  Following
the approach taken by Gehrels to improve his approximation to
$\lambda_l$ (see the following Section) we therefore apply a heuristic
correction to G86 Eq.~9 in the form of an additional term
$b\,(n+1)^c$:
\begin{equation}
 \lambda_u \approx (n+1)\; \left[ 1 - \frac{1}{9\,(n+1)} +
                   \frac{S}{3\,\sqrt{n+1}} + b\,(n+1)^c \right]^3 \;.
\end{equation}
We determine $b=b\,(S)$ from the requirement that the above
approximation be an identity for $n=0$, i.e., 
\begin{equation}
b\,(S) = \lambda_u\,(0,S)^{1/3}-1+1/9-S/3\;, 
\end{equation}
thus forcing better performance for low values of $n$. For each value
of $S$ from 0 to 7, $c=c\,(S)$ is then chosen such that the error of
the above approximation is minimized for $0\le n\le 100$. The runs of
$b$ and $c$ as functions of $S$ are shown in Figure~3. We find $c$ to
be negative for all values of $S$ with one-sided singularities at the
roots of Eq.~5, which lie at $S_{0,1}=0.50688$ (corresponding to a
confidence level of 69.388\%) and $S_{0,2}=2.27532$ (confidence level
98.856\%).

To allow the evaluation of Eq.~4 for any number of observed events,
$n$, and any confidence level, $S$, specifically in the proximity of
the mentioned singularities, we fit piecewise polynomial functions to
$b\,(S)$ and $c\,(1/S)$, or $c\,(\log_{\rm 10} S)$, with the degree of
the polynomial being determined by the requirement that the absolute
of the residuals be less than 1\% over the full $S$ range of the fit,
except at the locations of the mentioned singularities. We find
acceptable polynomial descriptions of $b\,(S)$ and $c\,(S)$ as
follows:
\begin{equation}
b\,(S) = \sum_{i=0}^{8} b_i\,S^i
\end{equation}
and
\begin{equation}
c\,(S) = \left\{ \begin{array}{l@{\quad:\quad}l} 
            \sum_{i=0}^{4} c_{1,i}\,\left(\frac{1}{S-S_{0,1}}\right)^i  & S<S_{0,1} \\
            \sum_{i=0}^{4} c_{2,i}\,\log_{10} (S-S_{0,1})^i & S_{0,1}<S<1.2 \\
            \sum_{i=0}^{3} c_{3,i}\,\left(\frac{1}{S-S_{0,2}}\right)^i & 1.2<S<S_{0,2} \\
            \sum_{i=0}^{7} c_{4,i}\,\log_{10} (S-S_{0,2})^i & S>S_{0,2} \\
               \end{array}\right.
\end{equation}
with $-10 \le c(S)\le 0$ overriding the above definition where
necessary, and coefficients $b_i, c_{1,i}, c_{2,i}, c_{3,i},$ and
$c_{4,i}$ as listed in Table~1. The results of these fits are shown as
the solid lines in Fig.~3.

Figure~4 shows the relative, absolute errors of Eq.~4 and demonstrates
that our approximation is accurate to better than 0.5\% for all values
of $n$ and $S$ considered here.

\section{Approximation of the Poissonian lower limit}
\label{lowlim}

As evidenced by Fig.~1 the Gaussian approximation
$\lambda_l(n,S)\approx n - S\sqrt{n}$ is a poor one for all but the
lowest values of $S$. G86 explores the behaviour of several more
sophisticated analytic approximations before resorting to modifying
the most promising of them by adding a heuristic power law term
$\beta\,n^\gamma$ (we used the same approach in the preceding section
to improve the approximation to $\lambda_u$):
\[
  \lambda_l \approx n\; \left( 1 - \frac{1}{9\,n} +
                   \frac{S}{3\,\sqrt{n}} + \beta\,n^\gamma \right)^3 \; .
  \hspace*{1cm} \mbox{(G86 Eq. 14)} 
\]
To find $\beta\,(S)$ and $\gamma\,(S)$ we proceed similarly as before
for our approximation to $\lambda_u$ and define $\beta\,(S)$ as
\begin{equation}
\beta\,(S) = \lambda_l\,(1,S)^{1/3}-1+1/9+S/3\;, 
\end{equation}
and then determine $\gamma\,(S)$ such that the error of the above
approximation is minimized for $0\le n\le 100$. The result, $\beta$
and $\gamma$ as functions of $S$, is shown in Figure~5 which, in the
overlap region, agrees with Fig.~1 of G86. $\gamma$ is negative for
all values of $S$ with a one-sided singularity at the only root of
Eq.~8 at $S_{0}=0.93876$, corresponding to a confidence level of
82.607\%.

To facilitate the evaluation of Eq.~14 of G86 for a wide range of
values of $n$ and $S$, we attempt to find analytical expressions for
$\beta\,(S)$ and $\gamma\,(S)$ (G86 quotes the values of either
function only at 10 locations between $S=1$ and $S=3.291$).  In
analogy to the approach taken in the preceding section, we fit
piecewise polynomial functions to $\beta\,(S)$, as well as to
$\gamma\,(S)$, $\gamma\,(1/S)$, or $\gamma\,(\log_{\rm 10} S)$, with
the degree of the polynomial being determined by the requirement that
the absolute of the residuals be less than 2\% over the full $S$ range
of the fit (less than 0.1\% at the high-$S$ end where high accuracy is
critical). We find acceptable polynomial descriptions of $\beta\,(S)$
and $\gamma\,(S)$ as follows:
\begin{equation}
\beta\,(S) = \left\{ \begin{array}{l@{\quad:\quad}l} 
            \sum_{i=0}^{5} \beta_{1,i}\,S^i & S\le 3 \\
            \sum_{i=0}^{4} \beta_{2,i}\,S^i & S>3\\
               \end{array}\right.
\end{equation}
and
\begin{equation}
\gamma\,(S) = \left\{ \begin{array}{l@{\quad:\quad}l} 
            \sum_{i=0}^{5} \gamma_{1,i}\,\log_{10} (S_0-S)^i & S\le S_{0} \\
            \sum_{i=0}^{5} \gamma_{2,i}\,\left(\frac{1}{S-S_0}\right)^i & S_{0}<S\le 2.7\\
            \sum_{i=0}^{2} \gamma_{3,i}\,S^i & S>2.7 \\
               \end{array}\right.
\end{equation}
with $-50\le\gamma\,(S)\le 0$ overriding the above where necessary,
and coefficients $\beta_{1,i}, \beta_{2,i}, \gamma_{1,i},
\gamma_{2,i},$ and $\gamma_{3,i}$ as listed in Table~2. The results of
these fits are shown as the solid lines in Fig.~5.

Equation 14 of G86 indeed yields greatly reduced errors when compared
to the exact values of $\lambda_l$. For $n<100$ and $1<S<3.291$ G86
quotes an accuracy of better than 2\% for the above
approximation. Figure~6 confirms this, but also demonstrates that the
errors become unacceptably large ($>10\%$) for higher confidence
levels and small to moderate values of $n$.

We now attempt to improve on Eq.~14 of G86 by adding a second,
higher-order correction term. As illustrated by Fig.~6 such an
additional term would have to improve the performance of the
approximation particulary in the high-$S$ regime for which Eq.~14 of
G86 was not optimized. This goal can be achieved by introducing a
(totally ad-hoc) sinusoidal term which adds only one additional
parameter $\delta$:
\begin{equation}
  \lambda_l \approx n\, \left[ 1 - \frac{1}{9\,n} +
            \frac{S}{3\,\sqrt{n}} + \beta\,n^\gamma + 
   \delta\,\sin \,\left(\frac{5}{n+1/4}\;\frac{\pi}{2} \right)\right]^3 \; .
\end{equation}
Since the sinusoidal term, by design, vanishes for $n=1$, Eq.~8 still
holds, and continues to define $\beta\,(S)$. $\gamma\,(S)$ (slightly
different from the one determined from Eq.~14 of G86) and
$\delta\,(S)$ are again obtained by iteratively minimizing the
absolute error of the approximation for $0\le n\le 100$. With
$\beta\,(S)$ unchanged and $\gamma\,(S)$ virtually indistinguishable
from the data shown in Fig.~5 we can focus on $\delta\,(S)$ which
shows a complex behaviour (Fig.~7). We do not attampt to model the run
of $\delta\,(S)$ for small values of $S$ where the function remains
close to zero. Instead, we fit a high-order polynomial to the high-$S$
end and set $\delta\,(S)=0$ for $S<1.2$:
\begin{equation}
\delta\,(S) = \left\{ \begin{array}{l@{\quad:\quad}l} 
                           0             & S<1.2 \\
            \sum_{i=0}^{8} \delta_i\,S^i & S\ge 1.2 \\
               \end{array}\right.
\end{equation}
with coefficients $\delta_i$ as listed in Table~3. The results of
these fits are shown as the solid line in Fig.~7.

Figure~8 demonstrates that Eq.~11 provides an approximation to
$\lambda_l$ that is accurate to better than 1\% when the polynomial
fits to $\beta\,(S)$, $\gamma\,(S)$, and $\delta\,(S)$ (Eqs.~9, 10,
12) are used, except for $n=1$ where an error of just over 1\% is
observed.

\section{Summary}

The Gaussian approximation $\lambda_{u} \approx n + S\sqrt{n}$ to the
true Poissonian upper confidence limit is acceptable for low
confidence levels ($S<3\sigma$) and $n>40$, but becomes increasingly
inaccurate for higher values of $S$. The situation is worse for the
Gaussian approximation $\lambda_{l} \approx n - S\sqrt{n}$ to the true
Poissonian lower confidence limit, which is off by more than 10\% at
$S=5$ even at $n=100$. The approximations proposed by G86 greatly
improve upon the Gaussian estimates but are still inaccurate at the
10\% level for low values of $n$ and high confidence levels.

Building on Gehrels' work we present improved algebraic approximations
which reduce the error with respect to the true Poissonian confidence
limits to under 1\% for $S\le 7$ (Poisson upper limit) and $S\le 5$
(Poisson lower limit). Although we have tested these equations only
for $n\le 100$, their analytic behaviour suggests that they hold for
all values of $n$ (cf.\ Figures 4 and 8).

To allow the numerical computation of approximate Poissonian
confidence limits for arbitrary combinations of $n$ and $S$ within the
quoted ranges, we provide the coefficients of piecewise polynomial
fits to all parameters used in the definition of either approximation.

All figures of this paper were produced using the Interactive Data
Language (IDL); the IDL source code of the approximations {\tt
poisson\_uplim} (Eq.~4) and {\tt poisson\_lolim} (Eq.~11) is available
from the author.\\

HE gratefully acknowledges financial support from NASA LTSA grant NAG
5-8253 and NASA ADP grant NAG 5-9238.

\begin{table}
\begin{tabular}{lccccc}
$i$ & $b_i$ & $c_{1,i}$ & $c_{2,i}$ & $c_{3,i}$ & $c_{4,i}$ \\
0 & $-3.8954$e$-03$ & $-2.0799$e$+00$ & $-1.4354$e$+00$ & $-8.4098$e$-01$ & $-1.0120$e$+00$  \\
1 & $+6.2328$e$-03$ & $-7.1925$e$-01$ & $-6.3188$e$-01$ & $+6.8766$e$-01$ & $-2.8853$e$-01$  \\
2 & $+5.2345$e$-03$ & $-4.0064$e$-01$ & $-1.6177$e$-01$ & $+2.0358$e$-01$ & $+4.2013$e$-01$  \\
3 & $-5.3096$e$-03$ & $-7.3386$e$-02$ & $-5.6966$e$-01$ & $+3.9965$e$-02$ & $-5.3310$e$-02$  \\
4 & $+1.3093$e$-03$ & $-5.4791$e$-03$ & $-2.2835$e$-01$ &                 & $-1.6319$e$-02$  \\
5 & $-2.0344$e$-04$ &                 &                 &                 & $+4.8667$e$-02$  \\
6 & $+2.0393$e$-05$ &                 &                 &                 & $-5.5299$e$-02$  \\
7 & $-1.1974$e$-06$ &                 &                 &                 & $-3.3361$e$-02$  \\
8 & $+3.1161$e$-08$ &                 &                 &                 &                  \\
\end{tabular}
\caption{Coefficients of polynomial fits to $b\,(S)$ and $c\,(S)$ of Eq.~4, as defined in Eqs.~6 and 7}
\end{table}

\begin{table}
\begin{tabular}{lccccc}
$i$ &   $\beta_{1,i}$   &    $\beta_{2,i}$    & $\gamma_{1,i}$ & $\gamma_{2,i}$ & $\gamma_{3,i}$ \\
0  & $-3.8605809$e$-03$ & $+3.4867327$e$-01$  & $-1.7480435$   & $-0.6347351$   & $-2.7517416$e$+00$  \\
1  & $-6.6002964$e$-03$ & $-4.0996949$e$-01$  & $-1.8895824$   & $-4.6707845$   & $+3.1692400$e$-01$  \\
2  & $+6.5798149$e$-03$ & $+1.6514495$e$-01$  & $-3.0808786$   & $+6.1602866$   & $-8.7788310$e$-03$  \\
3  & $+2.8172041$e$-03$ & $-1.5783156$e$-02$  & $-5.5164953$   & $-4.3543401$   &                     \\
4  & $+2.9892915$e$-03$ & $+5.2768918$e$-04$  & $-3.9940504$   & $+1.4470675$   &                     \\
5  & $-5.4387574$e$-04$ &                     & $-1.0248451$   & $-0.1870896$   &                     \\
\end{tabular}
\caption{Coefficients of polynomial fits to $\beta\,(S)$ and $\gamma\,(S)$ of G86, Eq.~14, as defined in Eqs.~9 and 10}
\end{table}

\begin{table}
\begin{tabular}{lcccccc}
%\rotate
$i$ &   $\beta_{1,i}$   &    $\beta_{2,i}$    & $\gamma_{1,i}$ & $\gamma_{2,i}$ & $\gamma_{3,i}$     & $\delta_i$ \\
0  & $-3.8605809$e$-03$ & $+3.4867327$e$-01$  & $-1.7174713$   & $-1.0131243$   & $-2.8115538$e$+00$ & $-2.2906640$e$-02$ \\
1  & $-6.6002964$e$-03$ & $-4.0996949$e$-01$  & $-1.7015942$   & $-2.9319339$   & $ 3.5117552$e$-01$ & $+6.8209168$e$-02$ \\
2  & $+6.5798149$e$-03$ & $+1.6514495$e$-01$  & $-1.9059468$   & $+3.2459998$   & $-1.3215426$e$-02$ & $-9.1678422$e$-02$ \\
3  & $+2.8172041$e$-03$ & $-1.5783156$e$-02$  & $-3.1324250$   & $-2.1348935$   &                    & $+7.1533924$e$-02$ \\
4  & $+2.9892915$e$-03$ & $+5.2768918$e$-04$  & $-2.0145052$   & $+0.6676902$   &                    & $-3.5010270$e$-02$ \\
5  & $-5.4387574$e$-04$ &                     & $-0.4257810$   & $-0.0834041$   &                    & $+1.0928872$e$-02$ \\
6  &                    &                     &                &                &                    & $-2.1069241$e$-03$ \\
7  &                    &                     &                &                &                    & $+2.2638722$e$-04$ \\
8  &                    &                     &                &                &                    & $-1.0302360$e$-05$ \\
\end{tabular}
\caption{Coefficients of polynomial fits to $\beta\,(S)$, $\gamma\,(S)$, and $\delta\,(S)$ of Eq.~11, as defined in Eqs.~9, 10, and 12}
\end{table}

\clearpage

\begin{figure}
\epsfxsize=0.8\textwidth
\hspace*{1cm}\epsffile{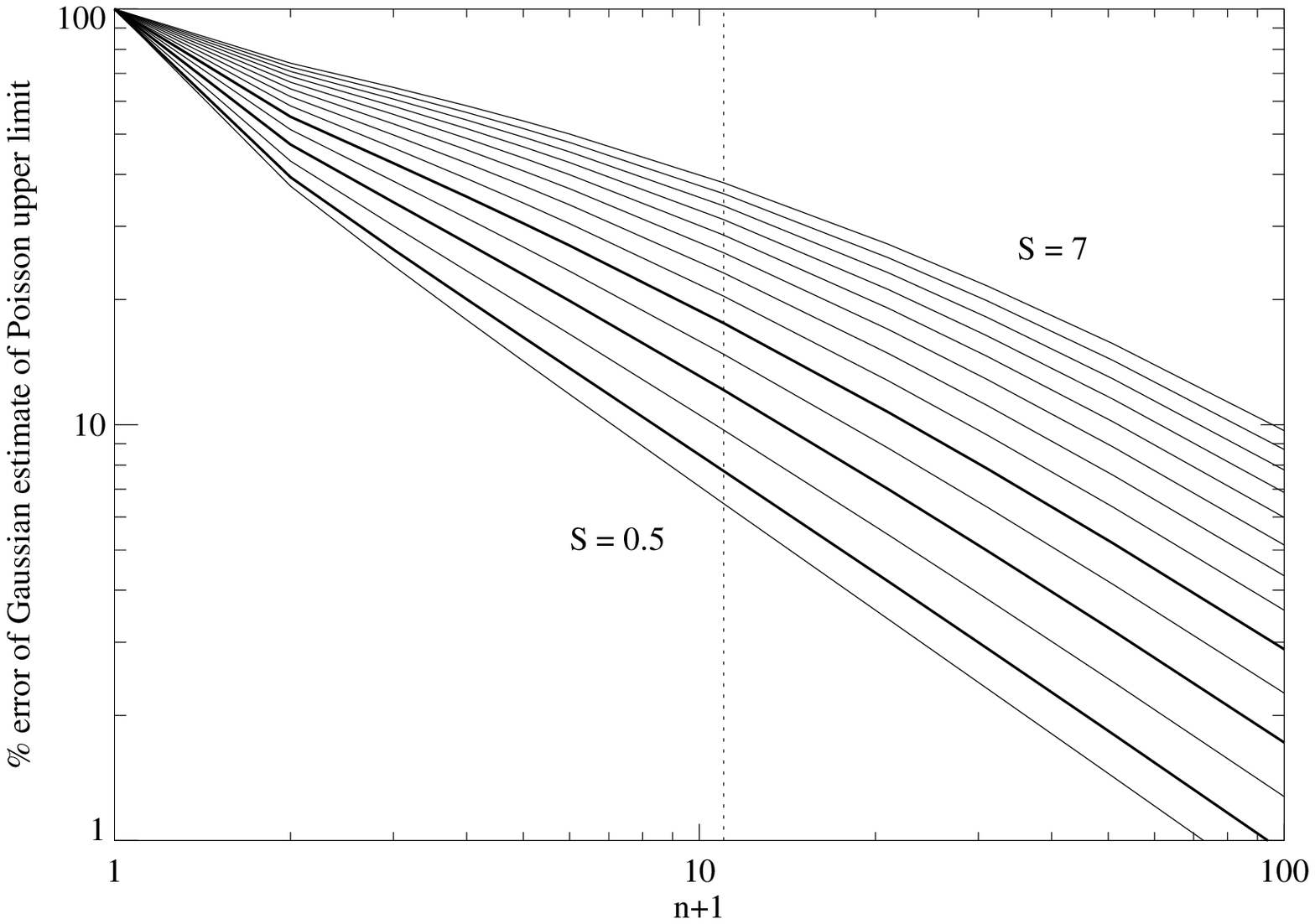}\\
\epsfxsize=0.8\textwidth
\hspace*{1cm}\epsffile{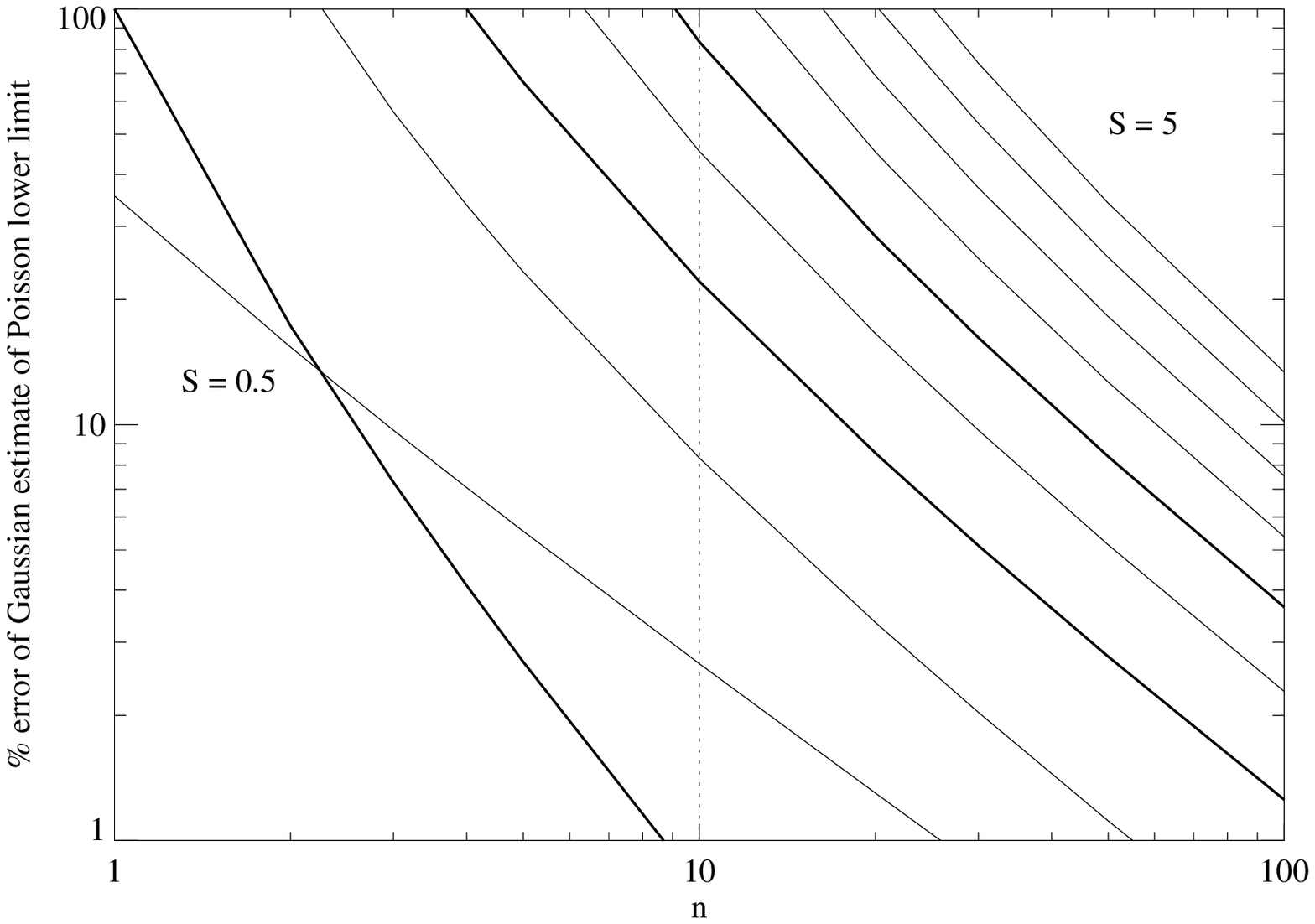}
\caption{Percentage error of the Gaussian approximations
$\lambda_u\approx n+S\,\sqrt{n}$ (top) and $\lambda_l\approx
n-S\,\sqrt{n}$ (bottom) as a function of n. In each panel the $S$
values of the shown curves vary from lower left to upper right in
steps of 0.5 as indicated; the thick lines correspond to $S=$1, 2, and
3. Note how, for $n=10$ (marked by the dotted line), the errors still
reach and exceed 10\% for all but the lowest values of $S$.}
\end{figure}

\clearpage

\begin{figure}
\epsfxsize=0.8\textwidth
\hspace*{1cm}\epsffile{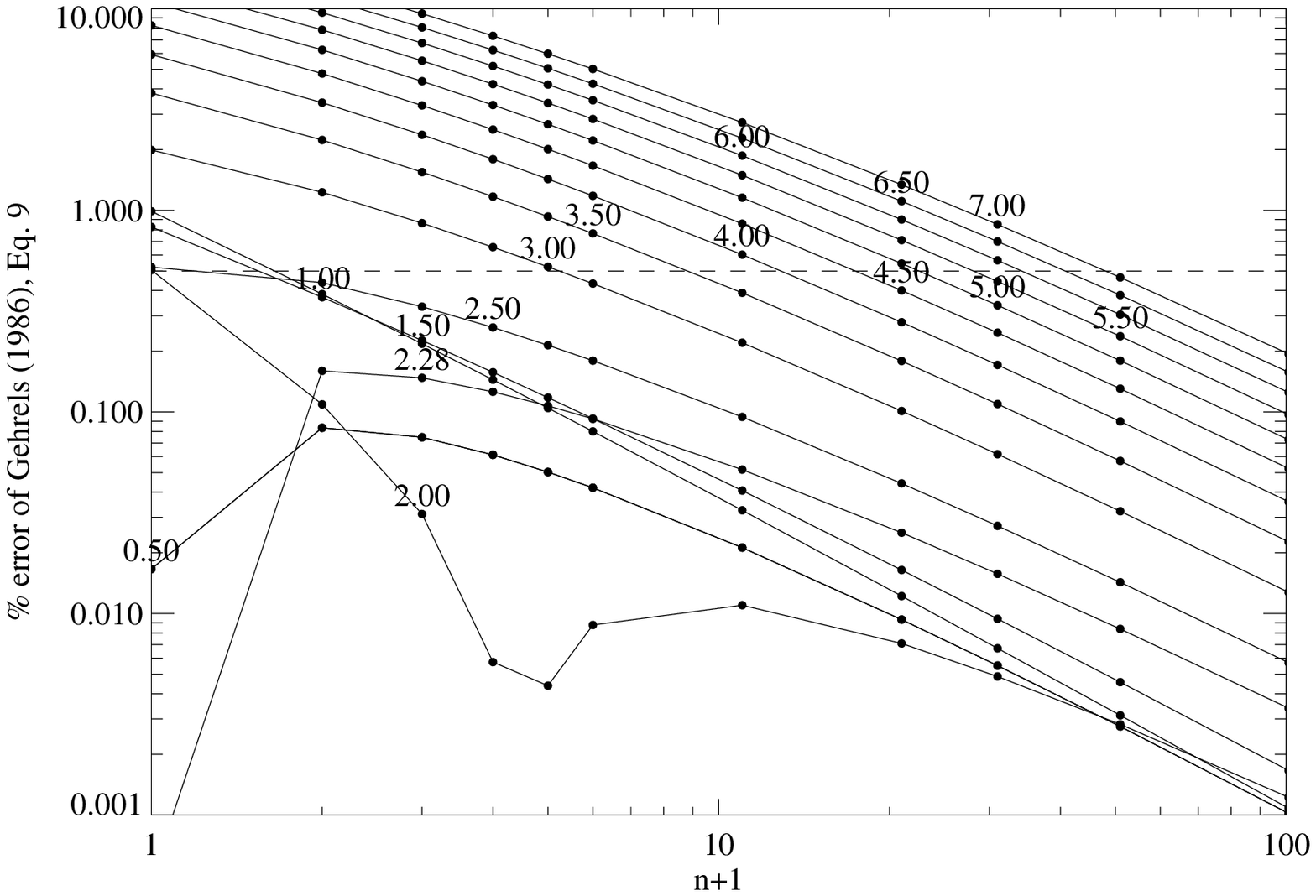}\\
\epsfxsize=0.8\textwidth
\hspace*{1cm}\epsffile{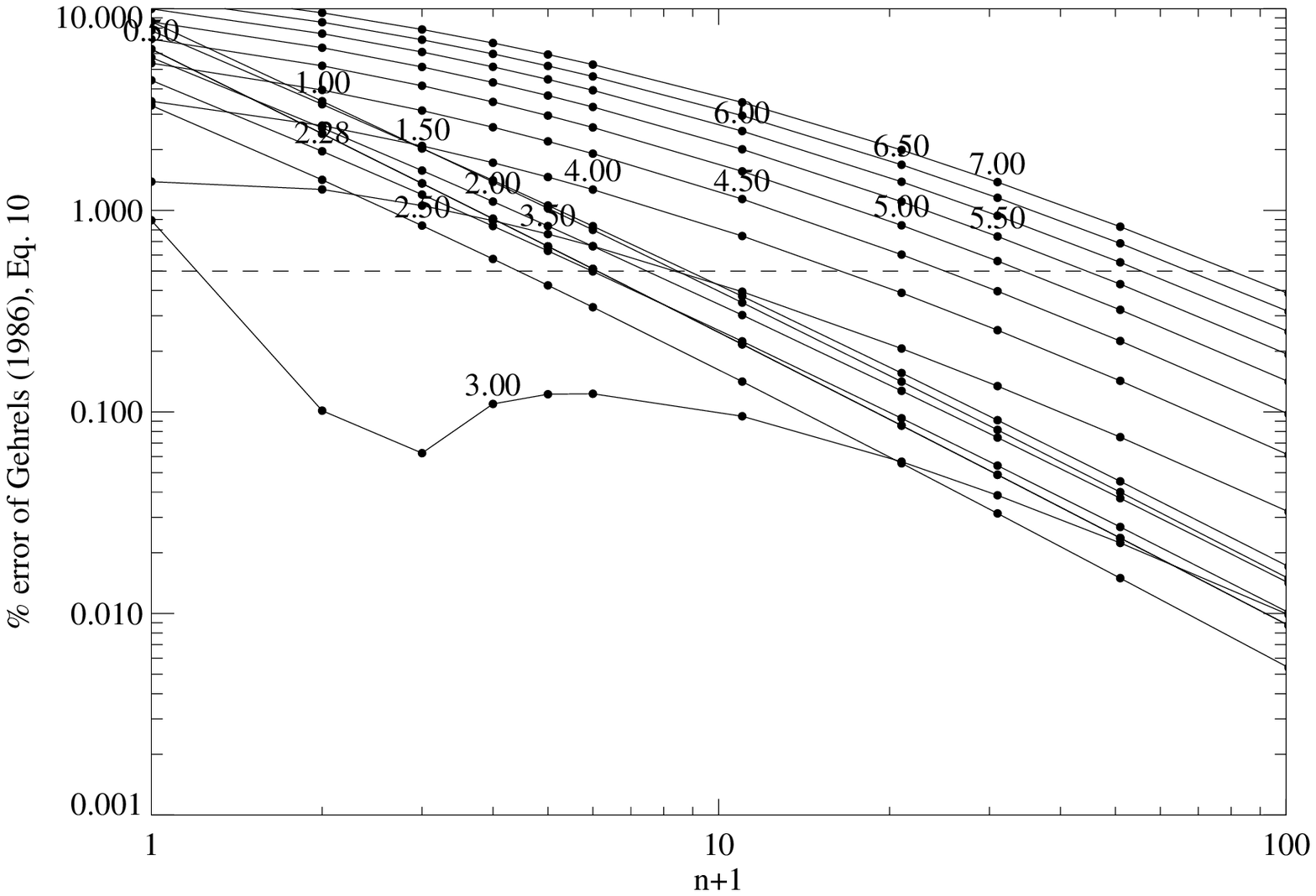}
\caption{Percentage error of the approximations of G86 Eq.~9 (top) and
G86 Eq.~10 (bottom) as a function of n. In each panel the $S$ values
of the shown curves vary from 1 to 7 as annotated. For essentially all
values of $n$ and $S$ explored here the error of either approximation
remains below the 10\% level for $n>2$, and below 1\% for
$n\ga 35$.  }
\end{figure}

\clearpage

\begin{figure}
\epsfxsize=0.8\textwidth
\hspace*{1cm}\epsffile{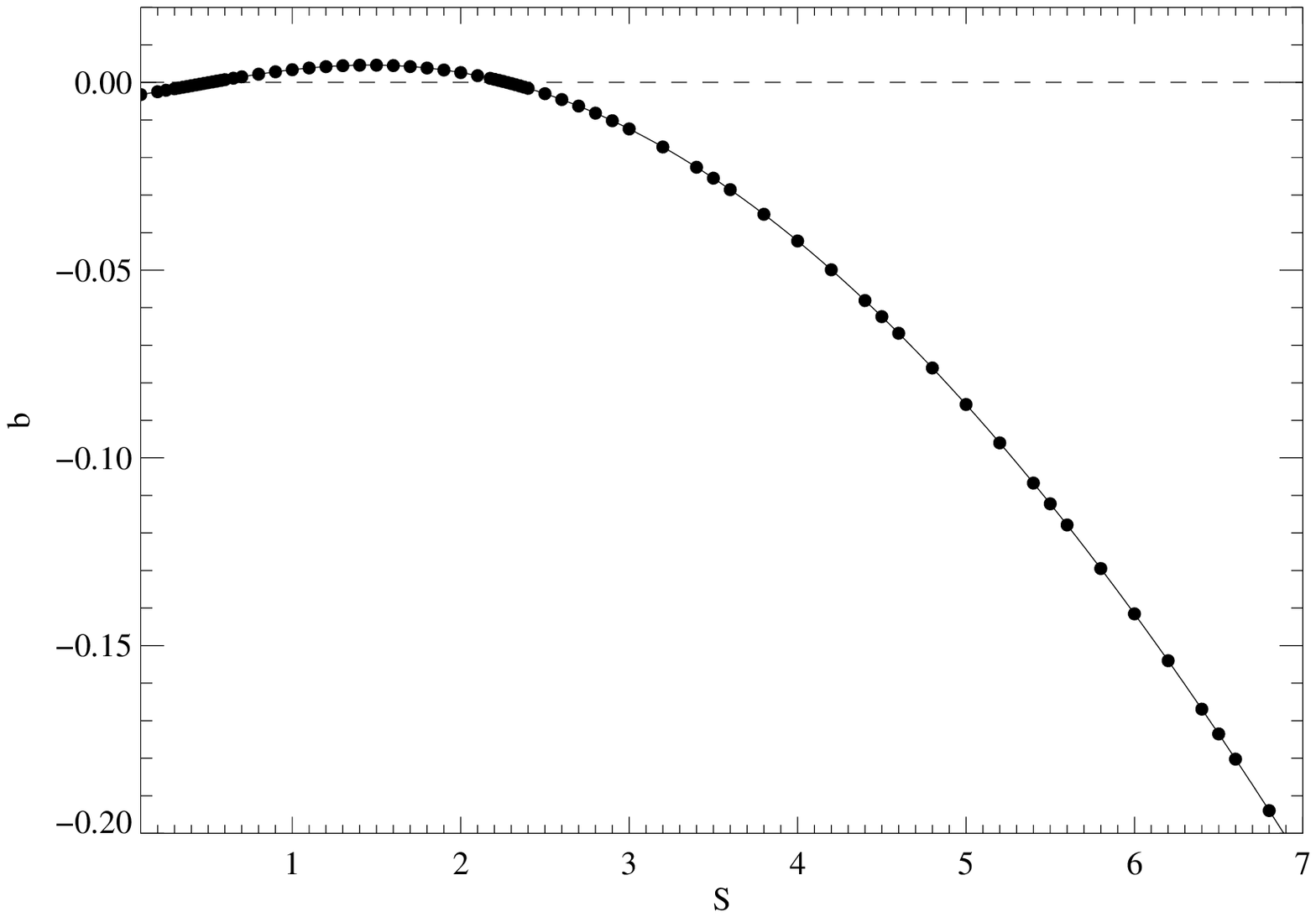}\\
\epsfxsize=0.8\textwidth
\hspace*{1cm}\epsffile{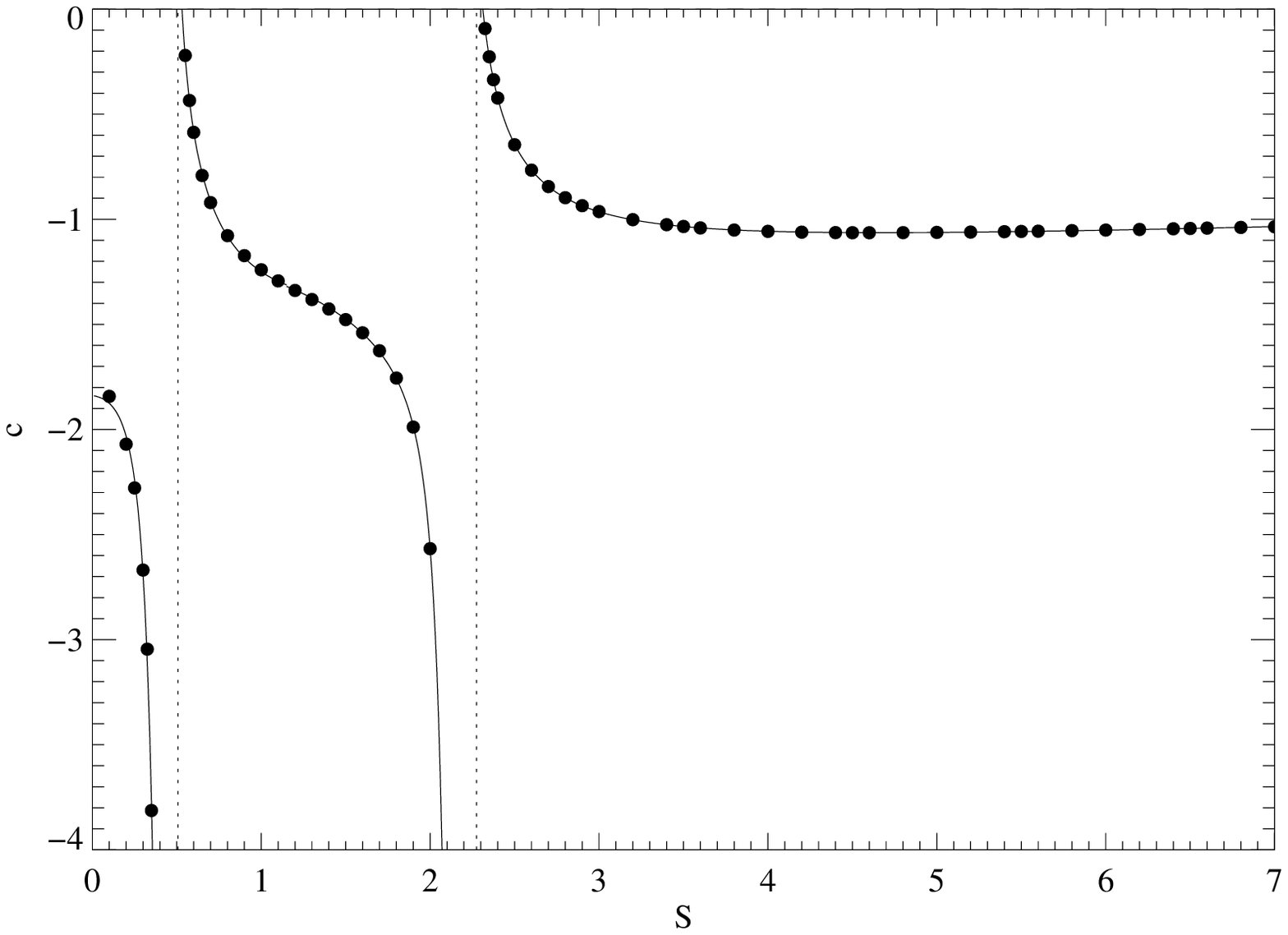}
\caption{Run of parameters $b$ (top) and $c$ (bottom) of Eq.~4 as a
function of $S$, the equivalent number of Gaussian $\sigma$. $c$
exhibits singularities at $S=0.507$ and $S=2.275$ where $b=0$. The
bullets mark the locations at which $b\,(S)$ and $c\,(S)$ were
computed, the solid lines mark polynomial fits to the data (see text
for details).}
\end{figure}

\clearpage

\begin{figure}
\epsfxsize=0.8\textwidth
\hspace*{1cm}\epsffile{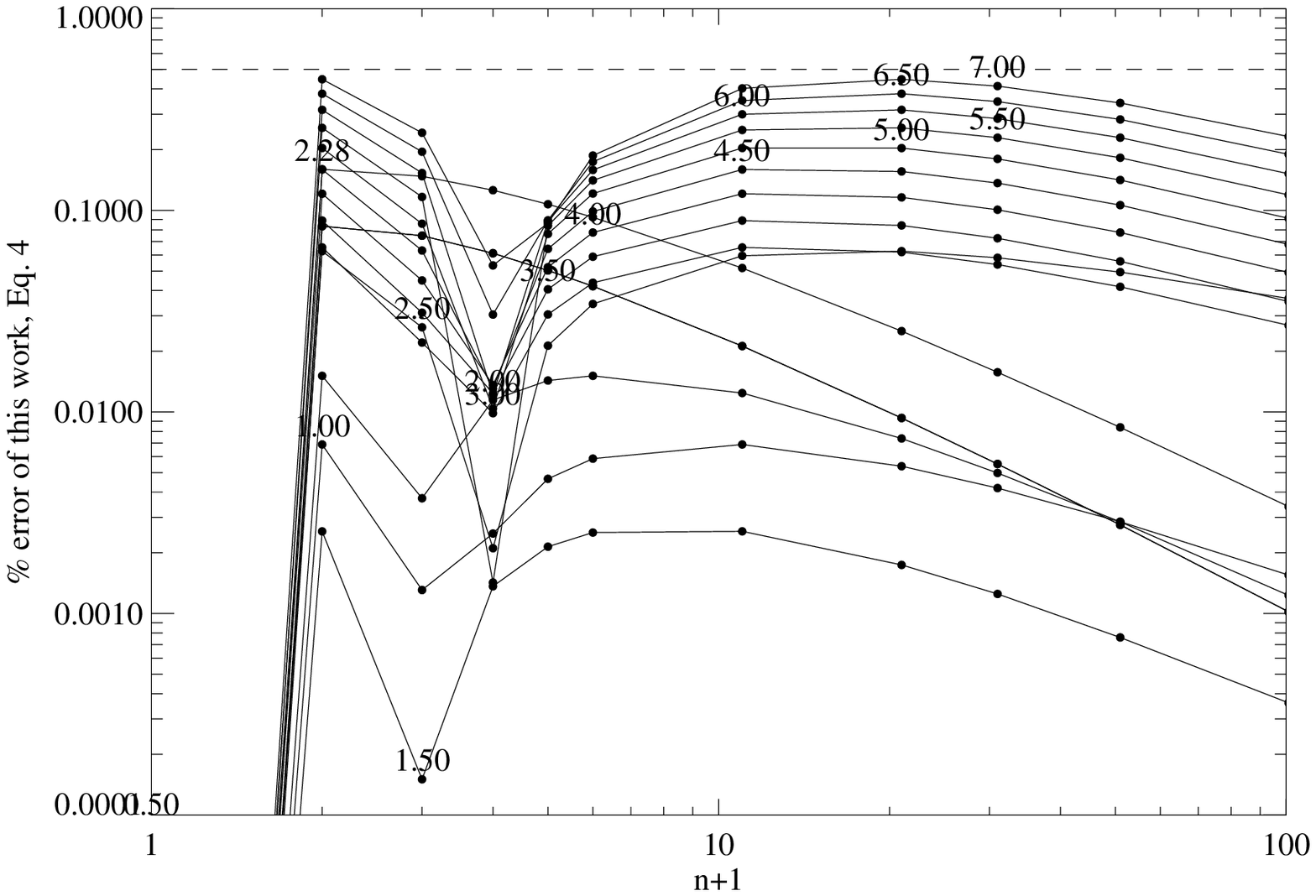}\\
\epsfxsize=0.8\textwidth
\hspace*{1cm}\epsffile{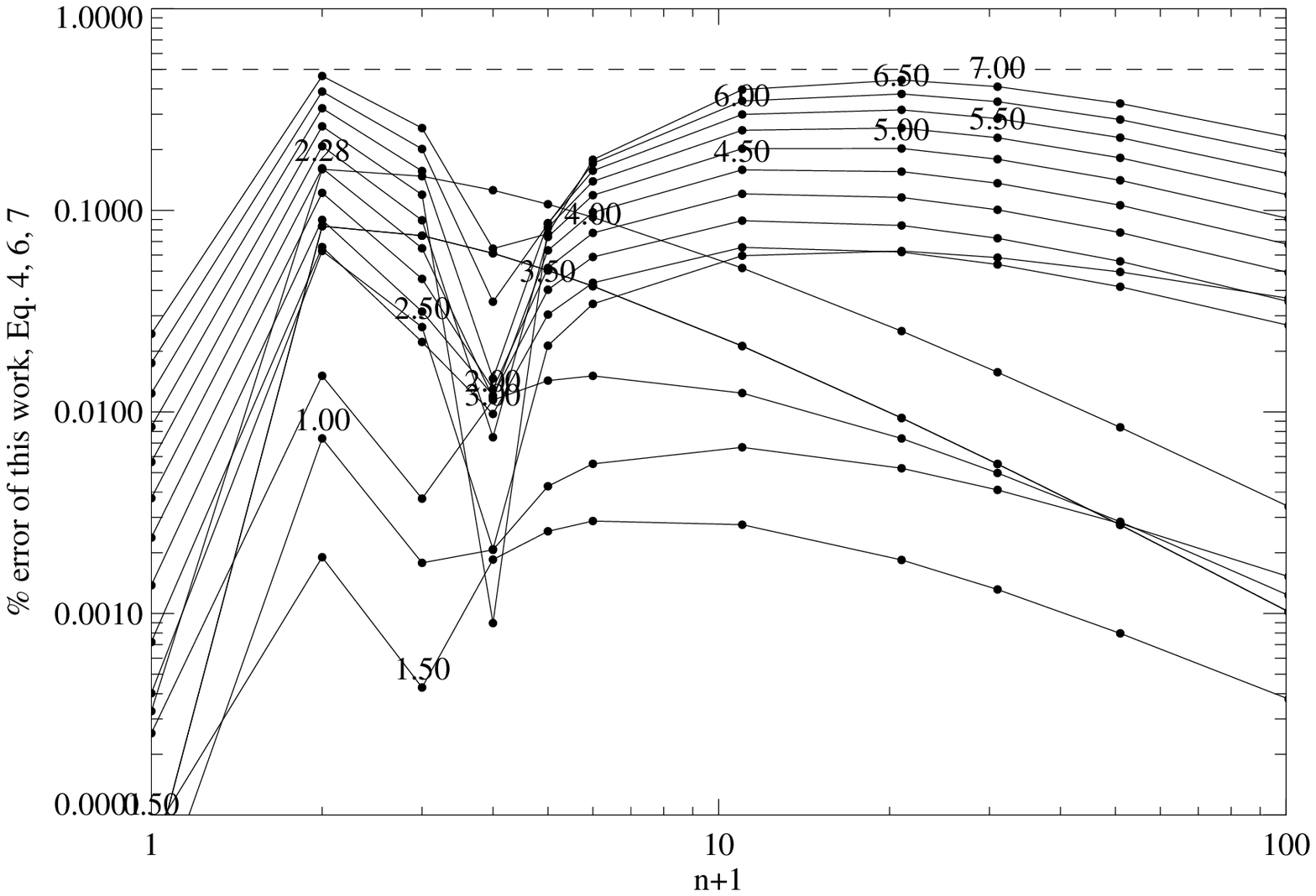}
\caption{Percentage error of the approximations of Eq.~4 with $b\,(S)$
and $c\,(S)$ as computed (top), and using the polynomial fits of Eqs.~6
and 7 (bottom), as a function of n. In each panel the $S$ values of the
shown curves vary from 0.5 to 7 as annotated. For all values of $n$
and $S$ explored here the error remains below the 0.5\% level.}
\end{figure}

\clearpage

\begin{figure}
\epsfxsize=0.8\textwidth
\hspace*{1cm}\epsffile{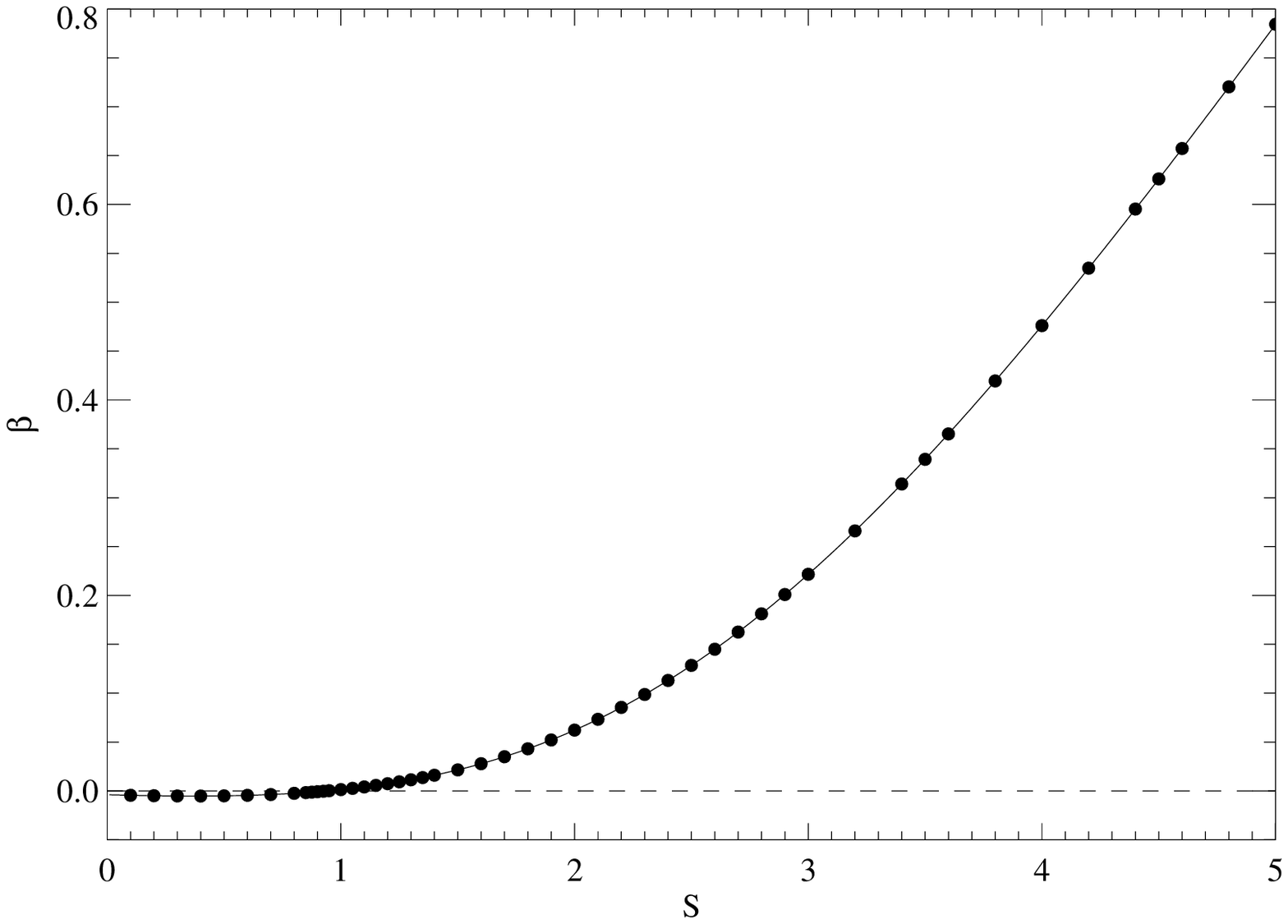}\\
\epsfxsize=0.8\textwidth
\hspace*{1cm}\epsffile{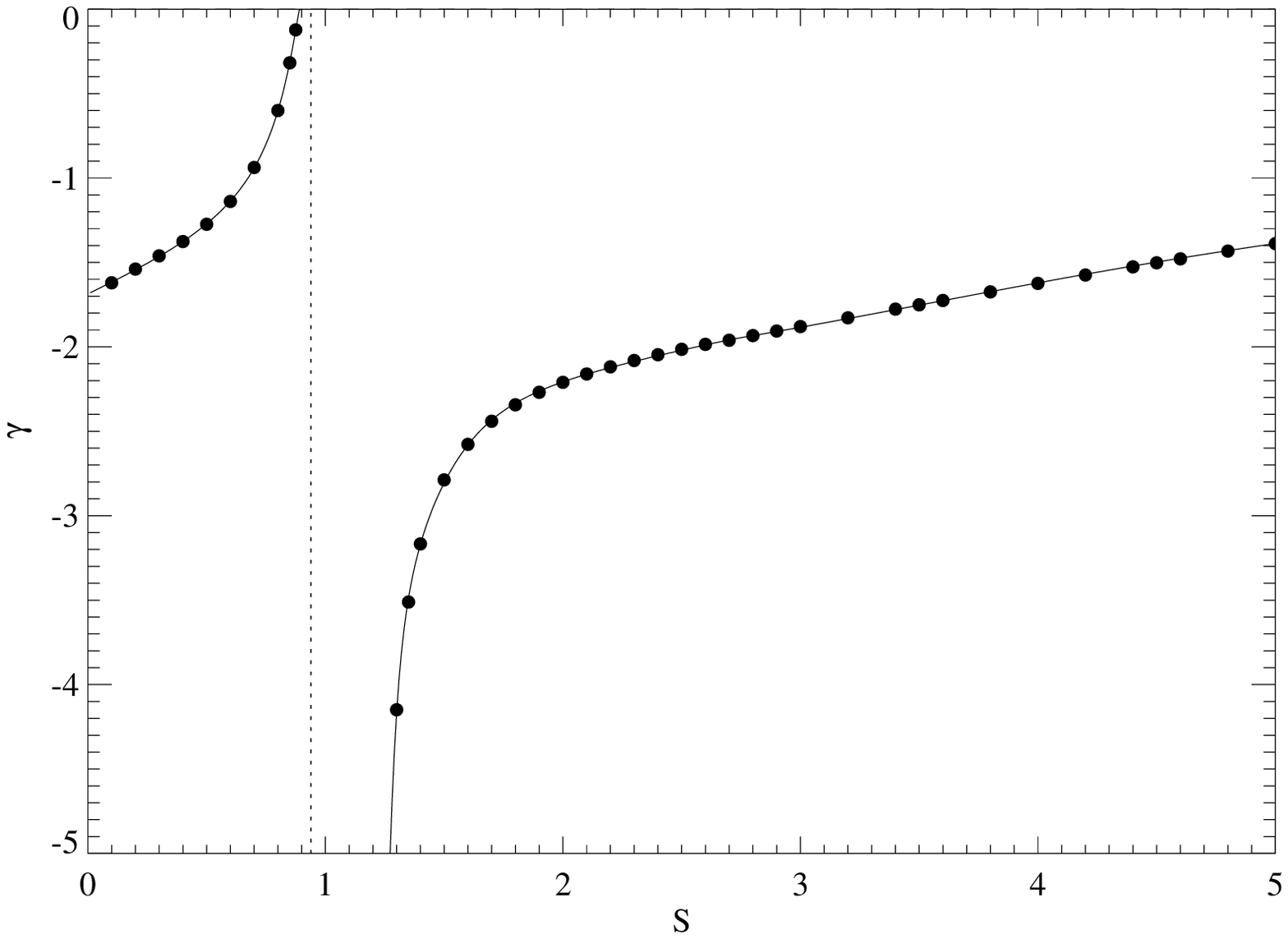}
\caption{Run of parameters $\beta$ (top) and $\gamma$ (bottom) of G86
Eq.~14 as a function of $S$, the equivalent number of Gaussian
$\sigma$. $\gamma$ exhibits a singularity at $S=0.939$ where
$\beta=0$. The bullets mark the locations at which $\beta\,(S)$ and
$\gamma\,(S)$ were computed, the solid lines mark polynomial fits to
the data (see text for details).}
\end{figure}

\clearpage

\begin{figure}
\epsfxsize=0.8\textwidth
\hspace*{1cm}\epsffile{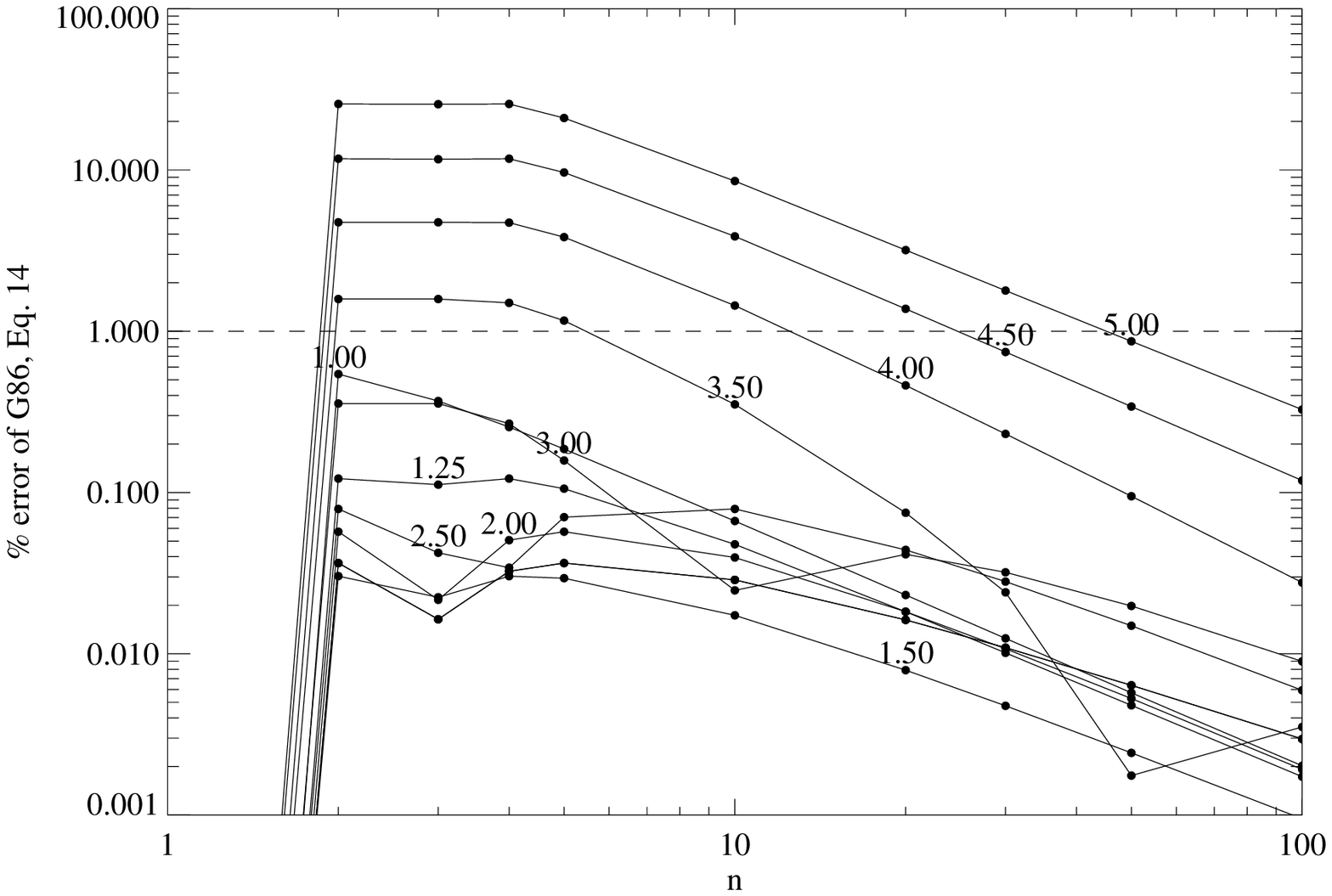}\\
\epsfxsize=0.8\textwidth
\hspace*{1cm}\epsffile{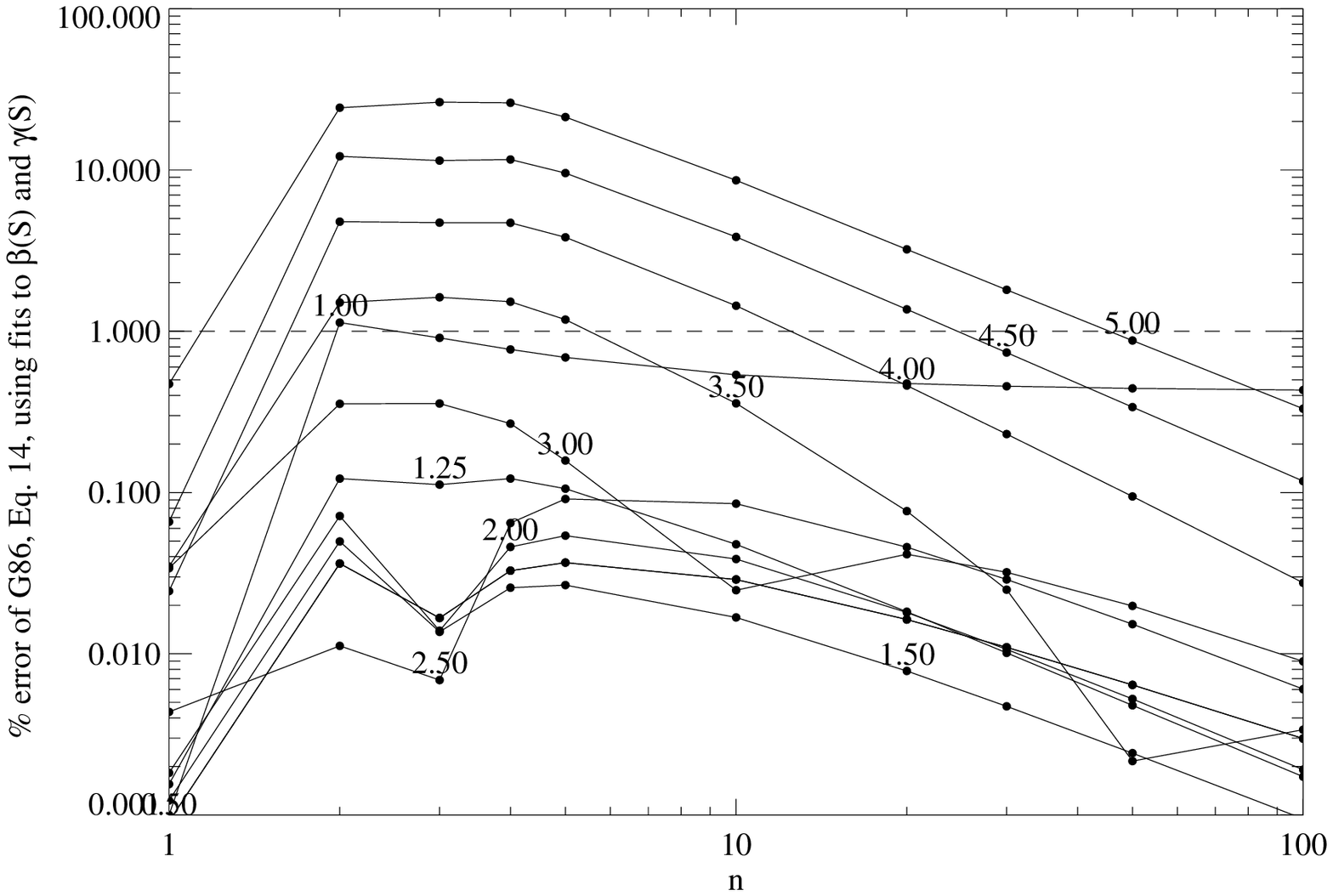}
\caption{Percentage error of the approximation to $\lambda_l$ given by
G86 Eq.~14 with $\beta\,(S)$ and $\gamma\,(S)$ as computed (top), and
using the polynomial fits of Eqs.~9 and 10 (bottom), as a function of
n. In each panel the $S$ values of the shown curves vary from 1 to 5
as annotated. While the approximation is good for low to moderate
confidence levels ($S<3.5$) it fails for $S>4$ where errors
approaching and exceeding 10\% are observed for small values of $n$.}
\end{figure}

\clearpage

\begin{figure}
\epsfxsize=0.8\textwidth
\hspace*{1cm}\epsffile{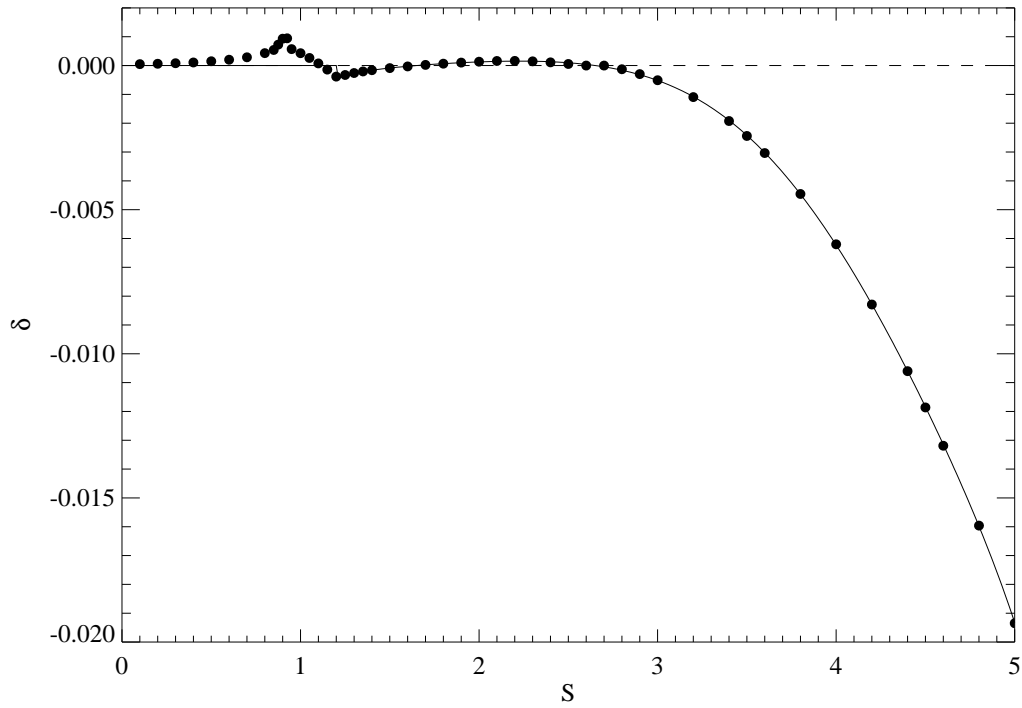}\\
\caption{Run of $\delta$ of Eq.~11 as a function of $S$, the
equivalent number of Gaussian $\sigma$. The bullets mark the locations
at which $\delta\,(S)$ was computed, the solid line marks a polynomial
fit to the data (see text for details).}
\end{figure}

\clearpage

\begin{figure}
\epsfxsize=0.8\textwidth
\hspace*{1cm}\epsffile{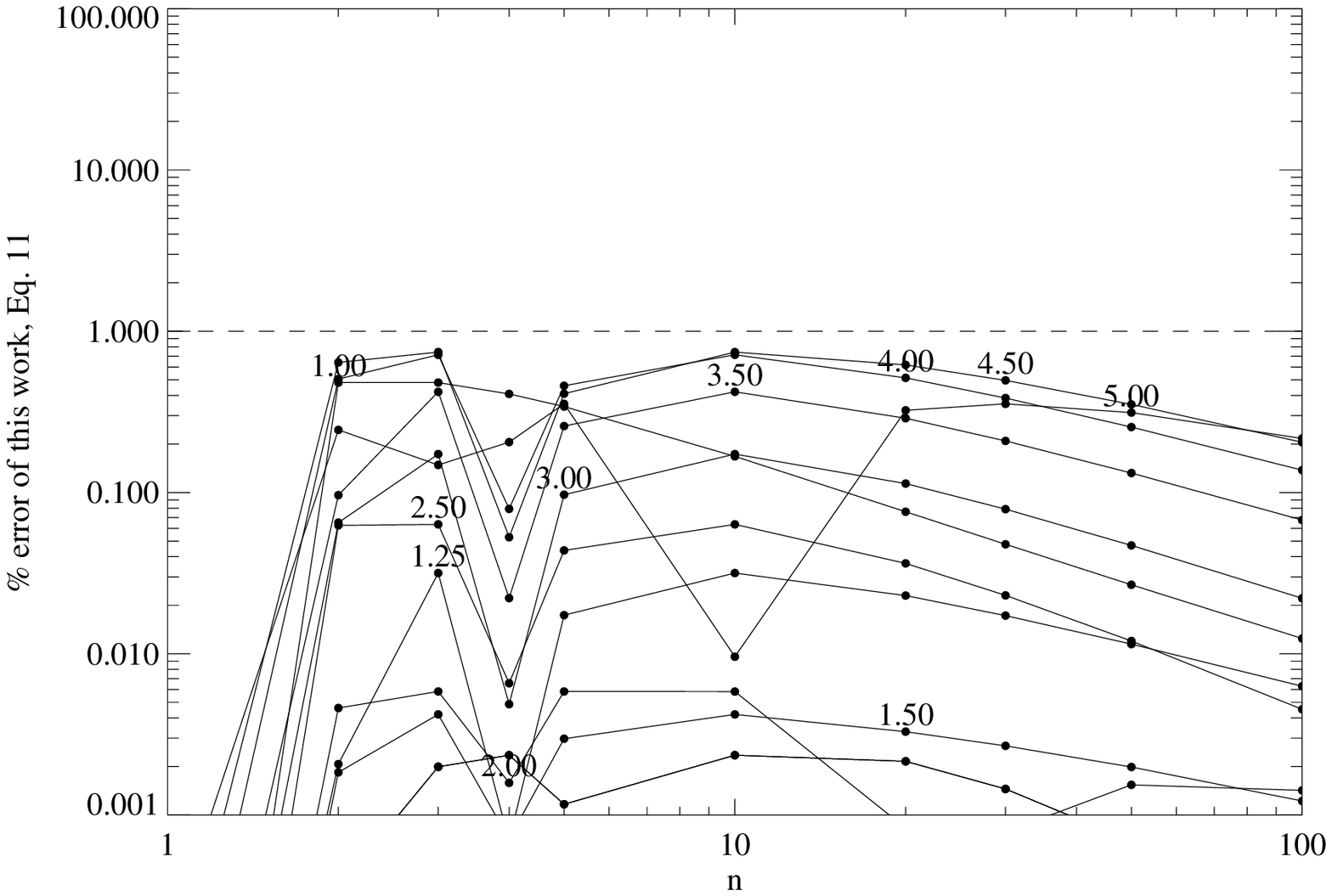}\\
\epsfxsize=0.8\textwidth
\hspace*{1cm}\epsffile{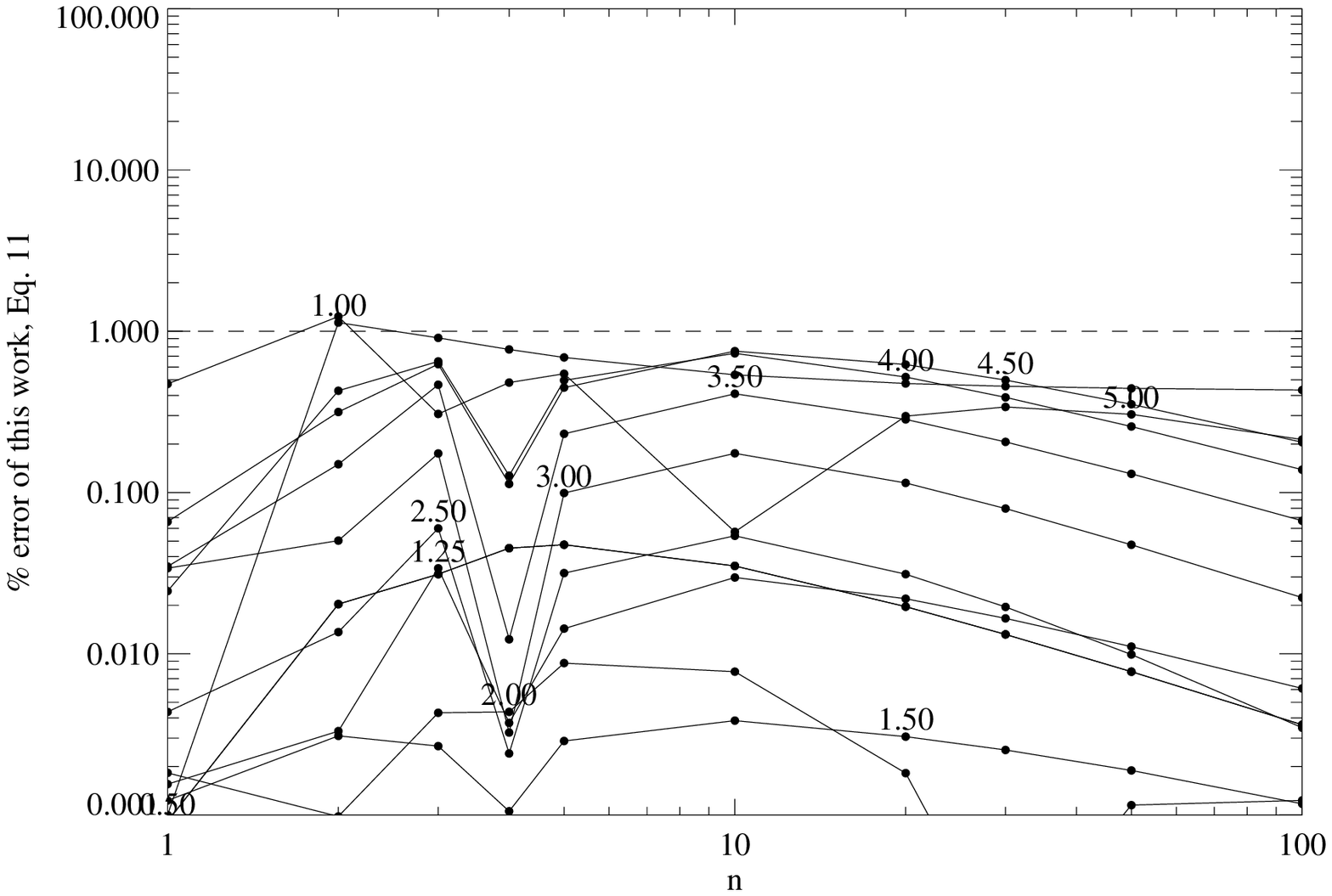}
\caption{Percentage error of the approximations of Eq.~11 with
$\beta\,(S)$, $\gamma\,(S)$, and $\delta\,(S)$ as computed (top), and
using the polynomial fits of Eqs.~9, 10, and 12 (bottom) as a function
of n. In each panel the $S$ values of the shown curves vary from 0.5
to 5 as annotated. For all values of $n$ (except $n=1$) and $S$
explored here the error remains below the 1\% level.}
\end{figure}

\end{document}